\documentclass[%
aps,pra,
reprint,
superscriptaddress,
twocolumn, 
nofootinbib,
floats,floatfix
]{revtex4-2}

\usepackage[english]{babel}

\usepackage[a4paper,top=2cm,bottom=2cm,left=2cm,right=2cm,marginparwidth=1.75cm]{geometry}

\usepackage{amssymb}
\usepackage{amsmath}
\usepackage{graphicx}
\usepackage{tabularx} % For better column width control
\usepackage{makecell} % For better column width control
\usepackage[colorlinks=true, allcolors=blue]{hyperref}

\begin{document}
\title{Simulating and Sampling from Quantum Circuits with 2D Tensor Networks}
\date{\today}

\author{Manuel S. Rudolph}
\email{manuel.rudolph@epfl.ch}
\affiliation{Institute of Physics, École Polytechnique Fédérale de Lausanne (EPFL),  Lausanne, Switzerland}
\affiliation{Centre for Quantum Science and Engineering, EPFL,    Lausanne, Switzerland}
\affiliation{Center for Computational Quantum Physics, Flatiron Institute, New York, New York 10010, USA}

\author{Joseph Tindall}
\email{jtindall@flatironinstitute.org}
\affiliation{Center for Computational Quantum Physics, Flatiron Institute, New York, New York 10010, USA}

\begin{abstract}
Classical simulations of quantum circuits play a vital role in the development of quantum computers and for taking the temperature of the field. Here, we classically simulate various physically-motivated circuits using 2D tensor network ans\"atze for the many-body wavefunction which match the geometry of the underlying quantum processor. We then employ a generalized version of the boundary Matrix Product State contraction algorithm to controllably generate samples from the resultant tensor network states. Our approach allows us to systematically converge both the quality of the final state and the samples drawn from it to the true distribution defined by the circuit, with GPU hardware providing us with significant speedups over CPU hardware. With these methods, we simulate the largest local unitary Jastrow ansatz circuit taken from recent IBM experiments to numerical precision. We also study a domain-wall quench in a two-dimensional discrete-time Heisenberg model on large heavy-hex and rotated square lattices, which reflect IBM's and Google's latest quantum processors respectively. We observe a rapid buildup of complex loop correlations on the Google Willow geometry which significantly impact the local properties of the system. Meanwhile, we find loop correlations build up extremely slowly on heavy-hex processors and have almost negligible impact on the local properties of the system, even at large circuit depths.
Our results underscore the role the geometry of the quantum processor plays in classical simulability. 
\end{abstract}

\maketitle

\section{Introduction}
Quantum circuits realize the non-equilibrium evolution of a many-body quantum system. Most familiarly, they are the ``programs" that quantum computers execute and, ideally, the measurement outcomes from such circuits provide the solution to some classically intractable, but useful problem.
\par
\begin{figure*}
    \centering
    \includegraphics[width=1\linewidth]{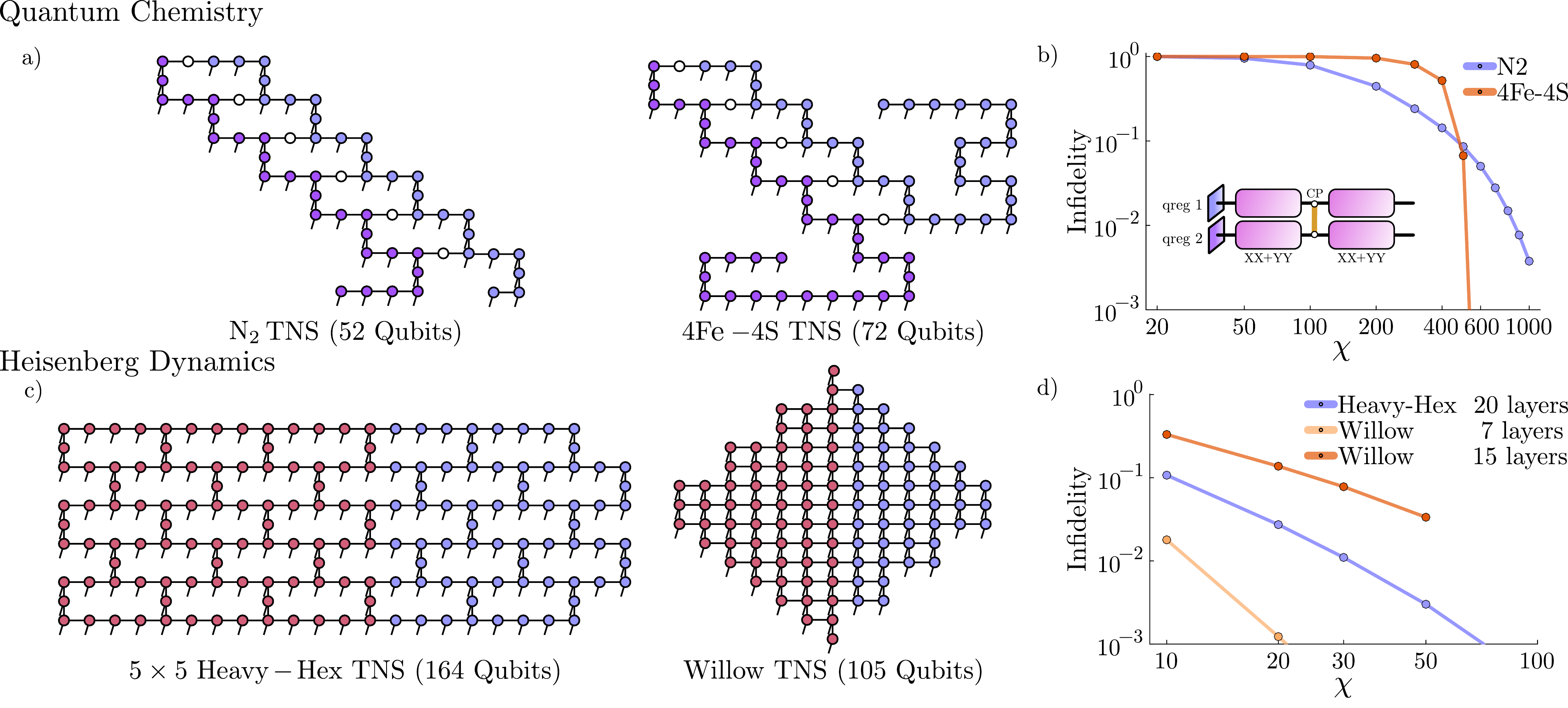}
    \caption{\textbf{Our tensor network topologies and their simulation errors.} 
    a) The tensor network states (TNS) for simulating the N$_{2}$ and $4$Fe-$4$S LUCJ circuits according to Ref.~\cite{robledomoreno2025}. Vertex colors indicate quantum sub-registers which encode the different spin states for the considered electronic system. b) Approximate infidelity (c.f. $1-f$ following Eq.~\eqref{eq:total_fidelity}) of the final TNS as a function of the maximum bond dimension $\chi$ in our simulations. We also sketch the circuit structures, which consist of single-site gates and XX+YY rotations within the quantum registers, and controlled phase (CP) gates between them.
    c) The TNS used for simulating discrete-time dynamics under the Heisenberg Hamiltonian on the heavy-hex or Willow processor topologies. Red and blue vertex colors indicate the regions of the model initialized in the 1- or the 0-state respectively. d) Approximate infidelity of the final TNS after simulating $20$ layers on the heavy-hex topology, and $7$ or $15$ on the Willow topology. Further details of the simulations can be found in Sec.~\ref{sec:results}.
    }
    \label{fig:fig1}
\end{figure*}

 Arguably the most prominent classical approach for simulating such circuits beyond the regime of exact diagonalisation is with the Matrix Product State (MPS)~\cite{Banuls2006, Yiqing2020, Dang2019optimisingmatrix, Napp2022EfficientClassical}, a one-dimensional flavor of \textit{tensor network state} (TNS). By tensor network state, we mean a general graph whose vertices consist of low-rank tensors and whose edges indicate along which tensor axes the wavefunction is factorized and entanglement is mediated (see Fig.~\ref{fig:fig1}). 
It is straightforward and efficient to extract information from states encoded as an MPS --- either via direct computation of the desired observable or by sampling bitstrings $x$ perfectly from the distribution of amplitudes $p(x) \sim  \vert \langle x \vert \psi \rangle \vert^{2}$ it encodes~\cite{stoudenmire2010minimally, ferris2012perfect}.

While MPS are extraordinarily effective for 1D and quasi-1D problems, many setups of interest follow more complex, higher-dimensional geometries. 
Prominently, quantum computers --- including the latest superconducting quantum processors --- typically involve qubits arranged in a planar lattice structure and with two-qubit gates applicable between neighboring qubits~\cite{arute2019quantum, kim2023evidence, jin2025observationtopologicalprethermalstrong}. As a result, even quantum states generated from fixed-depth quantum circuits on these architectures require resources growing exponentially with the number of qubits to capture with an MPS ansatz~\cite{kim2023evidence, king2025beyondclassical, haghshenas2025digital}. This is because of the effective mapping of a 2D state to a 1D ansatz, which results in the von-Neumann entanglement entropy on a bond of the MPS growing linearly with the height of the 2D system.

A more natural ansatz for the wavefunction of such systems is a tensor network whose geometry reflects that of the underlying processor and the interactions encoded in the quantum circuit~\cite{tindall2023efficient}.
Quantum states generated from \textit{fixed-depth} quantum circuits on the aforementioned 2D geometries are then guaranteed to be representable as a tensor network of fixed bond dimension with memory requirements that are only \textit{linear} in the number of qubits. The cost of extracting information from such networks and the dependence of the bond dimension with circuit depth is where the complexity of the problem presents itself. 
Despite their appealing nature as an ansatz, there is a relatively small body of work demonstrating the use of 2D tensor networks for simulating quantum circuits and benchmarking them against experimental data~\cite{tindall2023efficient, lee2025scalablesimulationrandomquantum, beguvsic2024fast, haghshenas2025digital}. Moreover, almost nothing is understood about how efficiently and accurately such networks can be sampled after application of a quantum circuit, a key component of many quantum algorithms.

Here we demonstrate the simulation and sampling of quantum circuits with 2D tensor networks in a controllable and verifiable manner. Our corresponding open-source software \cite{TensorNetworkQuantumSimulator} can be used to perform robust tensor network simulations of circuits realized over \textit{any} planar quantum processor, with GPU support enabling these samples and expectation values to be obtained with unprecedented speed. We consider two different circuits: \textbf{i)} the local unitary cluster Jastrow ansatz (LUCJ) circuits employed in Ref.~\cite{robledomoreno2025} and simulated on cutouts of IBM's latest processors with 52 and 72 qubits, and \textbf{ii)} the discrete-time dynamics of a domain wall quench of a 2D Heisenberg model on both IBM's heavy-hex architecture with 164 qubits and the latest Willow processor by Google with 105 qubits~\cite{abanin2025constructiveinterferenceedgequantum}. 

In all cases, we demonstrate how, for fixed depth, with systematically increasing computational resources we can obtain a tensor network representation of the state of the system with increasing fidelity and draw samples whose distribution converges to that exactly realized by the underlying circuit. Crucially, we identify efficiently computable metrics which attest to the quality of the state, expectation values, and bitstrings drawn from it.
In the case of the LUCJ circuits employed in Ref.~\cite{robledomoreno2025}, we find that the biggest system ($72$ qubits) can be simulated and sampled to numerical precision with our 2D tensor network approach. In general, we observe how the tree-like nature of the heavy-hex processors enables rapid and accurate tensor network simulations of deep quantum circuits with strikingly low levels of loop correlations.
Meanwhile, the denser grid structure of the Willow chip can give rise to challenging loop correlations even at relatively short circuit depths. Our results shed important light on the role of geometry in the classical simulability and physical nature of states realized by quantum circuits. 

\section{Methods}

\subsection*{Quantum Circuit Simulation}
We use a tensor network ansatz for the approximate state $\vert \psi_{N_{\rm G}} \rangle$ of the many-body wavefunction following the application of a quantum circuit $U = \prod_{i=1}^{N_{\rm G}} G_i$ consisting of a sequence of one- and two-qubit gates $G_{1}, G_{2} \hdots G_{N_{\rm G}}$ to an initial state $\vert \psi_{0} \rangle$. These tensor network states are a compressed format for the coefficients of the many-body wavefunction, and consist of a network of tensors --- one for each qubit --- connected by virtual indices which mediate the entanglement between the qubits in the system. The structure of the tensor network is chosen to reflect the geometry of the underlying quantum device upon which such a circuit might be implemented: with examples shown in Fig.~\ref{fig:fig1} including heavy-hex lattices mirroring IBM's current quantum devices and Google's latest Willow chip~\cite{abanin2025constructiveinterferenceedgequantum}. Whilst our focus here is on qubit-systems and one or two-qubit gates, extensions to more general qudit setups are straightforward\footnote{For the memory and time scalings given in this paper, we have assumed a local physical dimension $d= 2$ (i.e. a qubit system). More general local qudit spaces of dimension $d$ induce a scaling prefactor of $\mathcal{O}(d)$ in memory (assuming $d \ll \chi$) and, at worst, $\mathcal{O}(d^{2})$ in time for the methods described here.} and non-nearest-neighbor gates could be achieved using either SWAP gates or encoding the circuit in more general, long-range tensor network operators. 

The memory footprint of a tensor network can be determined simply from the size of the individual tensors and scales as $\mathcal{O}(N_{\rm qubits} \chi^{z})$, where $N_{\rm qubits}$ is the number of qubits (and thus tensors in the network) and $\chi$ is the maximum \textit{bond dimension} of any of the virtual indices in the network and $z$ is the \textit{coordination number}, i.e. the maximum number of virtual indices that any of the tensors possess. For heavy-hex geometries $z=3$ and for the Willow processor geometry $z = 4$.

One-qubit gates can be applied \textit{exactly} to the tensor network in $\mathcal{O}(\chi^{z})$ time without any truncation and do not alter the bond dimension. Meanwhile, two-qubit gates $G_{i}$ are applied \textit{exactly} to the state $\vert \psi_{i-1} \rangle$ and then the resulting combined tensor (consisting of the two tensors where the gate was applied and the gate itself) is truncated, via a Singular Value Decomposition (SVD), to a maximum dimension $\chi \leq \chi'$, where $\chi'$ is the exact bond dimension needed to keep all non-zero singular values. The procedure has time complexity $\mathcal{O}(\chi^{z + 1})$, and yields an approximate tensor network representation $\vert \psi_{i} \rangle \approx G_{i} \vert \psi_{i-1} \rangle$ of the state with equality occurring when $\chi = \chi'$. 

In this work, we apply two-qubit gates via a procedure that is mathematically equivalent to performing the well-known \textit{simple update} procedure from within the Vidal gauge~\cite{Vidal2003, Vidal2004, Jiang2008}.
The SVD is performed conditioned on a factorizable representation of the contraction of the network $\langle \psi \vert \psi \rangle$ surrounding the given two-qubits. This factorization takes the form of an outer product of ``message tensors", which can be obtained via the belief propagation (BP) algorithm~\cite{Alkabetz2021, tindall2023gauging} in $\mathcal{O}(N_{\rm qubits}\chi^{z+1} )$ time. An illustration is provided in Fig. S2 of the Supplementary Material (SM). The underlying approximation of this scheme, the BP approximation~\cite{tindall2023gauging}, is exact when the virtual indices of the tensor network do not form loops.
Even in the presence of loops, however, the true gate fidelity $\vert \langle \psi_{i} \vert G_{i} \vert \psi_{i-1}\rangle \vert^{2}$ is often well correlated with the sum of the square of the singular values $\sigma_{i}$ discarded~\cite{lee2025scalablesimulationrandomquantum}, allowing us to define an \textit{approximate gate error }
\begin{equation}\label{eq:error_per_gate}
    \epsilon_{i} =  \sum_{j = \chi + 1}^{\chi'}\sigma_{j}^{2} \approx 1 - \vert \langle \psi_{i} \vert G_{i} \vert \psi_{i-1}\rangle \vert^{2},
\end{equation}
where we have, for brevity, assumed the tensors are normalized such that $\sum_{j = 1}^{\chi'}\sigma_{j}^{2} = 1$. While in the presence of loops this is an approximation (it is exact when there are no loops~\cite{Yiqing2020}), increasing $\chi$ is still a reliable parameter which, in almost all practical cases, lowers $\epsilon_{i}$ and improves the accuracy of the tensor network representation. Moreover, there is the crucial guarantee that when the bond dimension is not truncated, i.e., $\epsilon_{i} = 0$, the tensor network is an exact representation of the many-body state, i.e. $\vert \psi_{i} \rangle = G_{i} \vert \psi_{i-1} \rangle$. It is useful to define the \textit{fidelity per gate} $f_i = 1 - \epsilon_i$, and from this an \textit{approximation} for the fidelity of the final state after application of the whole circuit
\begin{equation}\label{eq:total_fidelity}
    f = \prod_{i=1}^{N_{\rm G}} f_i\ \approx \vert \langle \psi_{N_{\rm G}} \vert \prod_{i=1}^{N_{\rm G}}G_{i} \vert \psi_{0} \rangle \vert^{2}.
\end{equation}
with the corresponding infidelity $1-f$.
Whilst this quantity is an approximation of the overall infidelity, in practical experience it is often a reliable error metric for the accuracy of the simulation in terms of overall state fidelity~\cite{Yiqing2020, lee2025scalablesimulationrandomquantum}.

When simulating entire circuits involving large numbers of gates, it is important that the message tensors remain updated so that the SVD truncations remain optimal under the BP approximation, and that $\epsilon_{i}$ is as accurate as possible. There is thus an important simulation cost trade-off to consider about the frequency with which BP should be re-run mid-circuit~\cite{tindall2023gauging} to update the message tensors. In our approach, we find that a sweet-spot is to update the message tensors between layers of non-overlapping gates. For structured circuits, such as those generated from the discrete-time dynamics of an underlying Hamiltonian, this naturally aligns with performing a Trotter decomposition and applying the two-qubit gates according to an \textit{edge coloring} of the underlying graph \cite{Richard1982EdgeColoring}. That is, we group the gates within one Trotter step into series of non-overlapping gates and re-run the BP algorithm between application of each series. As a result, the number of BP updates required during the circuit is independent of the system size and the overall time-complexity of our circuit simulation scales as $\mathcal{O}(N_{\rm qubits} \chi^{z+1} L)$ where $L$ is the number of (Trotter) steps.

\subsection*{The Boundary Matrix Product State Method for Planar Tensor Networks}
In this work, in order to extract information (via direct measurement of observables or via sampling) from the tensor network $|\psi\rangle$ following application of a circuit, we will need to contract both the \textit{norm network} $\langle \psi \vert \psi \rangle$ and \textit{amplitude networks} $\langle x \vert \psi \rangle$, where $\langle x|$ is a tensor network of bond dimension $\chi=1$ encoding a given bitstring $x$.

The BP algorithm can readily be used for fast, approximate contraction of these networks. In fact, we can introduce an approximate but efficiently computable BP error metric which is obtainable from the spectrum of eigenvalues of the transfer matrices formed using primitive loops (the set of $N_{l}$ loops of smallest size) of the tensor network. Specifically selecting a loop, inserting BP messages on the boundary and cutting open the virtual indices on a selected edge of the loop $l$ yields a matrix with eigenvalues  $\lambda^{l}_{1}, \lambda^{l}_{2}, \hdots $ sorted in decreasing order by their absolute value (see SM Fig. S3 for an illustration). We can then define errors representing a first-order approximation to the true BP error \cite{tindall2024confinement} in the network as 
\begin{align}
    \varepsilon &= \frac{1}{N_{l}}\sum_{l = 1}^{N_{l}}\varepsilon_{l} \label{eq:avBPError}\\
    \varepsilon_{l} &= 1 - \frac{\vert \lambda^{l}_{1} \vert}{\sum_{i} \vert \lambda^{l}_{i} \vert} \label{eq:loopBPerror} ,
\end{align}
where $\varepsilon$ is averaged over the per-loop BP errors $\varepsilon_l$. When calculated for $\langle \psi \vert \psi \rangle$, we have $0 \leq \varepsilon \leq 1 - \frac{1}{\chi^{2}}$ and this quantity is a very helpful indicator of the ``loop correlations" associated with the tensor network $\vert \psi \rangle$. 

It is necessary when $\varepsilon$ is large to go beyond BP, ideally with a systematically improvable approach, to contract the tensor network. This would allow convergable, more accurate information to be obtained whilst still avoiding the prohibitively high polynomial scaling with $\chi$ of near-exact contraction approaches.

In this work, we achieve this by adopting a \textit{boundary Matrix Product State} (MPS) contraction approach which is commonly used on open boundary square lattice tensor networks~\cite{verstraete2004renormalization, lubasch2014PEPS}. Notably, we have realized a more general implementation of the algorithm which works on any tensor network that, upon some grouping of the tensors into \textit{partitions} $\vert \psi_{b} \rangle$, $b = 1, 2, \hdots N_{b}$, where $N_b$ is the total number of partitions, forms a line. This means our contraction approach works on any planar tensor network, i.e. one that can be drawn in two dimensions without any edges crossing, such as those depicted in Fig. \ref{fig:fig1}. A typical partitioning that we will adopt is based on the columns of the networks depicted in Fig. \ref{fig:fig1} --- although partitioning based on diagonal or horizontal cuts is straightforward and supported in our codebase \cite{TensorNetworkQuantumSimulator}.

Following the partitioning, MPS of maximum virtual bond dimension $R$ can be passed through the partitions of the planar tensor network via sequentially fitting Matrix Product State - Matrix Product Operator (MPO) contractions. The resultant MPS form approximations for the partial contraction of the network with the approximation becoming equality in the limit $R \rightarrow \infty$. We have implemented an optimized, efficient MPS-MPO fitting algorithm and corresponding code which is highly general, in that the MPO can be a tensor network of \textit{any} structure which maps one MPS to another. Importantly, the one-site fitting procedure we adopt scales more favorably with bond dimension in comparison to density matrix or SVD-based contraction methods which have also recently been adapted to more general tensor network structures~\cite{ chubb2021generaltensornetworkdecoding, ma2024approximate} 

As a consequence, we have the automated ability to systematically contract any planar tensor network in a highly efficient, controllable manner. This allows us to controllably extract expectation values (including non-local correlators~\cite{tindall2025dynamics}) and, crucial to this work, sample bitstrings $x$ from $\vert \psi \rangle$ via an implementation of the TNS sampling procedure introduced in Ref.~\cite{Vieijra2021sampling}, but generalized to arbitrary planar topologies. See Fig. \ref{fig:fig1} for the specific geometries we consider in this work. Moreover, as our boundary MPS approach is dominated by tensor contractions and QR decompositions, significant speedups are observed when using GPU hardware --- which we will show and exploit explicitly in this work to rapidly and accurately contract 2D tensor networks of very high bond dimension.

\subsection*{Sampling from Tensor Network States}

\begin{table*}

\begin{tabularx}{\textwidth}{@{\extracolsep{\fill}} |c c c c c c|}  
\hline
 Processor & BP: $\langle \psi \vert \psi \rangle$ & \makecell{BMPS: $\langle \psi \vert \psi \rangle$ \\ ($R \leq \chi$)} & \makecell{BMPS: $\langle \psi \vert \psi \rangle$ \\ ($R > \chi$)} & \makecell{Sample $x \sim q(x)$} & \makecell{BMPS $\vert \langle x \vert \psi \rangle \vert^{2}$ \\ $R = \mathcal{O}(\chi)$} \\
 \hline
 Willow & $\mathcal{O}(\chi^{5})$ & $\mathcal{O}(R^{3}\chi^{5})$ & $\mathcal{O}(R^{2}\chi^{6}) + \mathcal{O}(R^{3}\chi^{4})$ & $\mathcal{O}(R^{3}\chi^{4})$ &  $\mathcal{O}(\chi^{6})$\\
 heavy-hexagonal & $\mathcal{O}(\chi^{4})$ & $\mathcal{O}(R^{3}\chi^{4})$ & $\mathcal{O}(R^{2}\chi^{5}) + \mathcal{O}(R^{3}\chi^{3}) $ & $\mathcal{O}(R^{3}\chi^{3})$ & $\mathcal{O}(\chi^{5})$
  \\
 \hline
\end{tabularx}
\caption{Scaling of the various contraction algorithms used in this work for tensor network states $\vert \psi \rangle$ of bond dimension $\chi$ and which match the Willow processor or heavy-hexagonal processor geometry. All algorithms scale linearly in $N_{\rm qubits}$, the number of tensors (one for each qubit). Boundary Matrix Product State (BMPS) contraction is performed using fitting (see SM Fig. S1 for more details), with MPSs of dimension $R$. The optimal choice of contraction sequences during fitting is dependent on whether $R \leq \chi$ or $R > \chi$. Sampling complexity is that of generating one sample from $q(x)$ (the BMPS approximation to $p(x) = \vert \langle x \vert \psi \rangle \vert^{2}$), given the pre-computed contraction of $\langle \psi \vert \psi \rangle$ using BMPS with MPS dimension $R$. This precomputed contraction can be re-used across samples. The final column refers to calculating $\vert \langle x \vert \psi \rangle \vert^{2}$ with BMPS, given a sample $x$ and using an MPS dimension $R = \mathcal{O}(\chi)$, which we find sufficient for obtaining accurate values of $\vert \langle x \vert \psi \rangle \vert^{2}$. This calculation is necessary if one wishes to verify the sample quality and compare the distribution $q(x)$ with $p(x)$. }
\label{Table:ComplexityTable}
\end{table*}

In the sampling procedure we distinguish between two probability distributions, $q(x)$ and $p(x)$, which are:
\begin{align}
    \nonumber q(x)&: \text{ the sampled distribution.} \\
    \nonumber  p(x)&: \text{ the actual distribution } \vert \langle x \vert \psi \rangle \vert^{2} \\&\text{\hspace{3mm} defined by the TNS}\nonumber.
\end{align}
The distribution we actually sample from, $q(x)$, is the one obtained by using boundary MPS of finite dimensions $R_{x}$ and $R_{n}$ to contract the networks $\langle x \vert \psi \rangle$ and $\langle \psi \vert \psi \rangle$ respectively. If we additionally find that, following the application of the circuit, the fidelity $f\approx1$, then $p(x)$ can be understood as the true distribution of the initial state evolved under the circuit.
We contract the norm network $\langle \psi \vert \psi \rangle$ once (independent of the number of samples) via MPS-MPO contractions in reverse order from partition $b = N_{b}, N_{b} - 1, \dots, 2$ and store those intermediate contractions. Then, for each sample, the partitions are sampled sequentially $b = 1, 2, \dots N_{b}$ with the network $\langle x \vert \psi \rangle$ contracted ``on-the-fly" as the partitions are moved through. More details and an in-depth illustration of the sampling procedure are provided in the SM (see, specifically, Fig.~\ref{fig:tns-sampling}). 

Importantly, we also calculate the ratio $p(x)/q(x)$, which attests to the quality of each sample. Whilst this can be computed ``on-the-fly" when sampling the partitions~\cite{Vieijra2021sampling}, this estimate is only accurate when a sufficiently large sampling MPS dimension $R_{x}$ is used. Instead, in this work, we allow ourselves to perform the sampling with arbitrary $R_{x}$ and $R_{n}$ and independently verify the samples by performing an accurate computation of $p(x) = \vert \langle x \vert \psi \rangle \vert ^{2}$ upon generation of the sample. Whilst this requires a separate tensor network contraction, it can be done more efficiently in comparison to using a large $R_{x}$ within the sampling procedure \textit{and} allows us freedom in choosing $R_{x}$ and $R_{n}$.

The mean probability ratio has the useful property that it can be used as an estimator of the norm, 
\begin{align}\label{eq:poverq_norm}
    \mathbb{E}_{x\sim q}\left[\frac{p(x)}{q(x)}\right]  &= \sum_x q(x) \frac{p(x)}{q(x)} \notag \\
    &= \sum_x p(x) = \langle\psi|\psi \rangle,
\end{align}
which is not necessarily $1$ for our tensor networks due to the truncations performed during the circuit application.
We emphasize this norm estimate is only unbiased if the support of $q(x)$ and $p(x)$ coincide. Additionally, even if it is unbiased, if there are cases where $q(x) \ll p(x)$  then it can be a poor estimate requiring an extremely large number of samples.

The $p(x)/q(x)$ ratio is an informative metric for assessing the quality of individual samples, but we compute one further metric that communicates the quality of the samples, the \textit{sample KL-Divergence} (KLD)~\cite{kullback1951information, Vieijra2021sampling}. The sample KLD reads
\begin{align}\label{eq:kld}
    \text{KLD}(q, p) &= \sum_x q(x) \log{\frac{q(x)}{p(x)}} \notag \\
    &= \mathbb{E}_{x\sim q}\left[\log{\frac{q(x)}{p(x)}}\right] ,
\end{align}
which is the (inverse) log-ratio of the probabilities averaged over the samples drawn from $q(x)$. Note that $p(x)$ formally needs to be normalized by the sample-based normalization in Eq.~\eqref{eq:poverq_norm}. A KLD of $0$ guarantees that the distributions are identical, but small values significantly below $1$ typically indicate high-quality samples. Conveniently, we can also estimate the stochastic uncertainty $\Delta$KLD via 
\begin{align}
    \Delta\text{KLD} = \text{STD}_{x\sim q}\left[\log{\frac{q(x)}{p(x)}}\right].
\end{align}

In this work, we will sample from tensor networks with heavy-hex or Willow geometries following the application of several different circuits. We will use identical MPS dimensions $R_{x} = R$ and $R_{n} = R$ when contracting $\langle x \vert \psi \rangle$ and $\langle \psi \vert \psi \rangle$ and certify our samples independently by contracting the network $\langle x \vert \psi \rangle$ via MPS-MPO contractions with an MPS of maximum bond dimension $2\chi$ (which we find sufficiently large to accurately compute $p(x)$ in all cases). Just like applying gates, the complexity of generating a number of samples $m$ is dependent on the coordination number $z$ of the tensor network. In Table \ref{Table:ComplexityTable} we list the time complexity scalings of the contraction approaches used in this work for the Willow and Heavy-Hex geometries.  We will explicitly see some of these scalings manifest in our data. It should be emphasized that the lower coordination number of the heavy-hex typically reduces the scaling of all algorithms by a factor of the state dimension $\chi$.

\section{Results}\label{sec:results}

\subsection*{Precise simulation of local unitary Jastrow ansatz circuits}

\begin{figure*}
    \centering
    \includegraphics[width=1\linewidth]{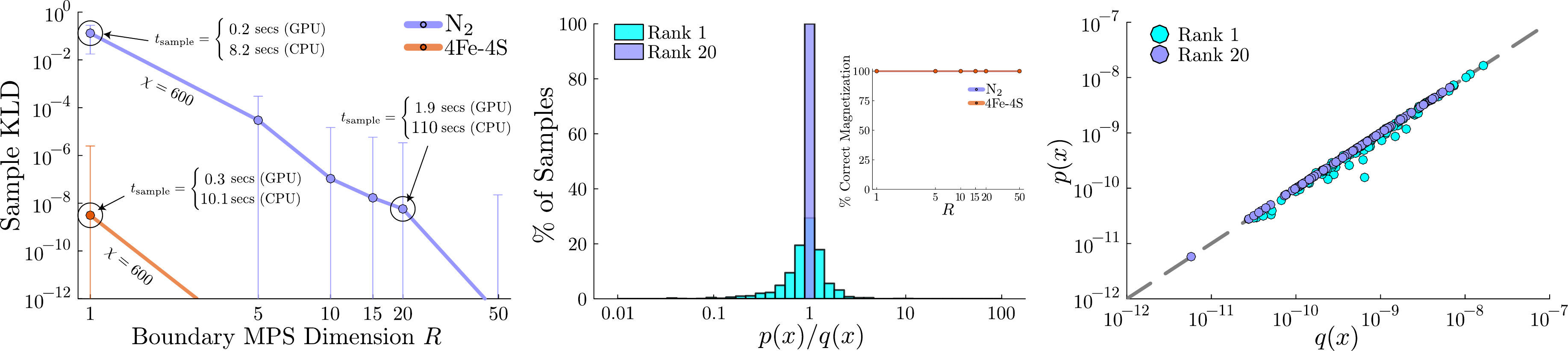}
    \caption{\textbf{Simulation and sampling of local unitary Jastrow circuits.} 
    \textbf{(left)} KLD (see Eq.~\eqref{eq:kld}) of $1000$ samples generated with boundary MPS of varying bond dimensions $R$ from a TN state with bond dimension $\chi=600$ after application of the corresponding LUCJ circuit. Error bars show the standard error on the KLD estimate.
    Annotated times indicate the average time $t_{\text{sample}}$ to generate a single bitstring using either an Intel Xeon Gold (CPU) or an Nvidia RTX A6000 (GPU), following a pre-computed contraction of the norm network $\langle \psi \vert \psi \rangle$ for a given $R$. The 72-qubit simulation of the $4$Fe-$4$S molecule exhibits effectively no loop correlations and can be sampled very accurately with $R=1$. \textbf{(middle)} Histogram of the $p(x)/q(x)$ probability ratios at sampling dimensions $R=1$ and $R=20$ for the N$_{2}$ molecule, confirming that at $R=20$ we generate practically exact samples. The inset showcases that samples generated at any $R$ lie within the correct magnetization subspace. 
    \textbf{(right)} $q(x)$ vs $p(x)$ scatter plot for $R=1$ and $R=20$.}
    
    \label{fig:chemistry_sampling}
\end{figure*}

We first turn our attention to the Local Unitary Cluster Jastrow ansatz (LUCJ) circuits employed in Ref.~\cite{robledomoreno2025} on IBM's processors. These circuits were used as part of an intricate hybrid framework where samples from the evolved states were used to diagonalize molecular Hamiltonians and find low-energy eigenstates. A potential quantum advantage within this framework could arise from classical simulations being unable to capture the evolved states with reasonable resources or from an inability to accurately sample from them in reasonable time.
In this work, using the tensor networks illustrated in Fig. \ref{fig:fig1}, we execute those same circuits and then generate numerically exact samples from the evolved states. 

We study the two largest circuits executed in Ref.~\cite{robledomoreno2025} for the N$_{2}$ molecule with 52 qubits and the $4$Fe-$4$S molecule with 72 qubits. The executed circuits can be found as part of that work, and consist of particle number preserving rotations - most prominently CP gates and ${\rm XX}+ {\rm YY}$ rotations. In Ref.~\cite{robledomoreno2025}, the ${\rm XX}+ {\rm YY}$ gates are transpiled into device-native CNOT gates, with the largest circuit (for the $4$Fe-$4$S molecule) involving $\sim  3500$ CNOTs. Here, we do not need such transpilation and thus run the same circuit in a format involving $\sim 1800$ ${\rm XX}+ {\rm YY}$ rotations. 
The circuit topologies are sublattices of IBM's heavy-hex processors with 6 and 4 primitive loops, respectively (see Fig.~\ref{fig:fig1}). Here, we show that, with a tensor network graph adapted to the circuit topology, we can simulate the 52-qubit case near-exactly and the 72-qubit case to numerical precision, with a time-to-sample far below one second on a single GPU. Note that Ref.~\cite{robledomoreno2025} reports 58 and 77-qubit simulations, where ancilla qubits had to be used to fit the problem into the heavy-hex topology of the quantum device. Here such ancilla are not necessary and can be viewed as just the presence of redundant identity matrices on the bonds of the tensor network (see white nodes in Fig. \ref{fig:fig1}).

In Fig.~\ref{fig:fig1} we show that we can achieve overall state fidelities (to three decimal places) of $f  = 0.996 \approx \vert \langle \psi \vert C \vert \psi_{0} \rangle \vert^{2}$ and $f = 1.000 \approx \vert \langle \psi \vert C \vert \psi_{0} \rangle \vert^{2}$ based on the discarded singular values during the circuit at $\chi=1000$ for the N$_{2}$ and $4$Fe-$4$S circuits $C$, respectively with $\vert \psi_{0} \rangle$ the initial product state and $\vert \psi \rangle$ the final state encoded in the tensor network. These map to mean CNOT gate fidelities of $99.9998 \%$ and $99.99999 \%$ respectively, which can be contrasted with the $99.8 \%$ reported in Ref.~\cite{robledomoreno2025}. 

Fig.~\ref{fig:chemistry_sampling} shows our results for sampling from the $\chi=600$ tensor network states, which are still of very high quality with $f = 0.950$ and $f = 1.000$ respectively. We refer the reader to Fig.~\ref{fig:all_chemistry_poverqs} in the SM for $p(x)/q(x)$ histograms for a wider range of $R$ values. Strikingly, even with the lowest sampling MPS dimension of $R =  1$ (recall we set $R_{x}, R_{n} = R$), i.e., using effectively a belief propagation approximation when generating samples, all generated samples have the correct magnetization despite no efforts being made to enforce the underlying U(1) conservation in the tensors in the network. Moreover, the sample KLD from Eq.~\eqref{eq:kld} --- which rigorously quantifies the sample errors --- is zero, to machine precision (using 64-bit arithmetic), for both problems when  $R = 50$. Empirically, we find that for N$_{2}$, KLD values of below $\sim 10^{-3}$ are achievable with $R = 5$ whilst for $4$Fe-$4$S below $\sim 10^{-8}$ is achievable with  $R = 1$ suggesting an almost total absence of loop correlations despite the high depth of the circuits. This appears to be a general pattern of heavy-hex topologies where loop correlations are strikingly small (even when accounting for the loop size) \cite{tindall2023efficient} compared to more dense 2D lattices. In this application, however, a significant culprit is also the low number of gates between the sub-registers (see the coloring in Fig. \ref{fig:fig1}). There are one or two CP gates per sub-register connection, which results in the inter-register bonds being of low dimension and entanglement along these bonds is directly responsible for the presence or absence of loop correlations in this system. In effect, due to the geometry of the problem, these systems can be simulated with two weakly coupled Matrix Product States, which is what our simulation approach achieves naturally, by virtue of its generality.

It is an interesting avenue of future research to consider LUCJ circuits which can generate states which are harder to classically simulate and / or extract information from. This could be either due to them having larger loop correlations or having more substantial entanglement. The two are often closely related. This complexity could be achieved by circuits which \textbf{i)} have a higher frequency of gates between the two sub-registers, \textbf{ii)} are implemented on a device with smaller loops or \textbf{iii)} come from CCSD (coupled cluster singles and doubles) calculations where the orbital rotation parameters generate substantially more entanglement. We consider this a pertinent avenue of future research.

\subsection*{Discrete-time dynamics of the Heisenberg model}
The Heisenberg model is a paradigmatic spin model which gives rise to rich quantum dynamics~\cite{Bertini2016XXZ, Ljubotina2017heisenbergdynamics}. 
The Hamiltonian reads 
\begin{align}\label{eq:Heisenberg}
    H = J\sum_{\langle i,j \rangle} & (X_iX_j + Y_iY_j + Z_iZ_j) = \sum_{\langle i,j \rangle} H_{ij}
\end{align}
where the summation runs over the neighboring sites of lattices, which is specified by the topology or connectivity of the system, and $X_{i},Y_{i},Z_{i}$ are the usual Pauli operators for the $i$th qubit. Here, we study real, discrete-time dynamics under this Hamiltonian until time $t$ according to a first-order Trotter-Suzuki decomposition of the evolution with discrete time steps $\delta t = t/L$ and $L$ layers. The decomposition used is based on an edge coloring of the lattice into a minimal number of groups of pairs of sites $E_{1}, E_{2}, \hdots E_{K}$ such that each site $i$ appears at most \textit{once} in a given group. 

The corresponding quantum circuit is
\begin{align}
U &= \prod_{l=1}^L U_l, \qquad U_l = \prod_{k = 1}^{K}\prod_{\langle i,j \rangle \in E_{k}} e^{-{\rm i}H_{ij} \delta t}
\label{eq:HeisenbergProp}
\end{align}
with the rotation matrices in a given group commuting with each other. We perform the necessary edge coloring for this decomposition via an implementation of the Misra and Gries Algorithm \cite{Misra1992}, which --- due to König's theorem \cite{konig1990} --- guarantees $K = z$ when the lattice is bipartite with $z$ the coordination number. On the heavy-hex lattice $z=3$ whilst for the Willow topology $z=4$. Both are bipartite lattices. For non-bipartite lattices, Vizing's theorem \cite{Vizing1964} guarantees $K \leq z+1$.

\begin{figure*}
    \centering
    \includegraphics[width=1\linewidth]{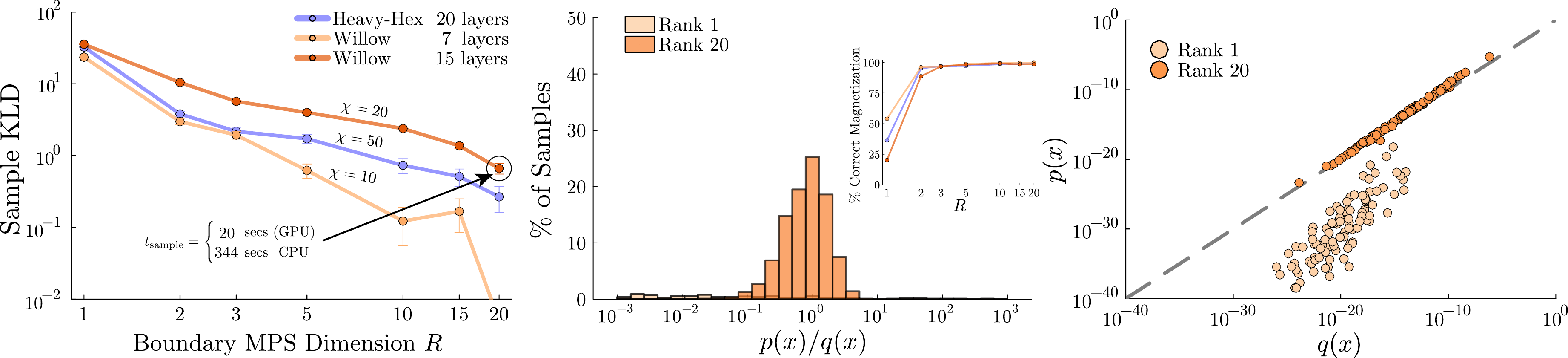}
    \caption{\textbf{Simulation and sampling of the quench dynamics of a domain wall for the Heisenberg model on heavy-hex and Willow topologies.} 
    \textbf{(left)} KLD (see Eq.~\eqref{eq:kld}) of $1000$ samples generated with boundary MPS with indicated bond dimensions $R$. Error bars show the standard error on the KLD estimate. Annotated time indicates the average time $t_{\text{sample}}$ (rounded up) to generate a single bitstring using either an Intel Xeon Gold (CPU) or an Nvidia RTX A6000 (GPU). 
    \textbf{(middle)} Histogram of the $p(x)/q(x)$ probability ratios at $R=1$ and $R=20$ for the more challenging $L=15$ Willow case. We find that at $R=1$ the probabilities of the samples are often off by dozens of orders of magnitude, which is why the bars within the shown limits are barely visible. In contrast, at $R=20$ we generate higher-quality samples with more concentrated probability ratios. The inset shows the percentage of samples that lie within the correct magnetization subspace, which is over $96\%$ in all cases at $R=3$. 
    \textbf{(right)} $q(x)$ vs $p(x)$ scatter plot, highlighting that the probabilities we sample with the rank $R$ approximation are frequently larger than they physically are in the encoded state.}
    \label{fig:heisenberg-sampling}
\end{figure*}

We draw inspiration from Ref.~\cite{Rosenberg2024}, which studied the magnetization transfer under Heisenberg evolution in a 46 qubit chain, where one half of the system favored initialization to the 0-state and the other half to the 1-state. The use of large time steps $\delta t$ allowed the quantum device based on superconducting qubits to reach entangled regimes whilst keeping the depth of the circuit reasonable compared to when using a smaller time step.
Here, we port these experiments to a 164-qubit heavy-hex topology with $5\times5 = 25$ primitive loops and the 105-qubit Willow chip topology, which are shown in Fig.~\ref{fig:fig1}c. We split the system into two halves and initialize them, as indicated in red and blue, in the 1-state and 0-states.

We set a time step of $\delta t=0.1$ and $J=1$. Because the individual rotation gates are $U(1)$, they preserve the total magnetization and the evolved states $|\psi\rangle = U|\psi_0\rangle$ are a superposition of basis states that each have the same number of 0s or 1s as the initial state $|\psi_0\rangle$. This has the benefit that --- alongside the fidelity bound and the sample KL divergence --- it serves as another metric for assessing the quality of our samples. As the states evolve, the initial domain wall becomes a continuous transition with local expectation values going from $Z_i\in\{-1, 1\}$ to $Z_i\in[-1, 1]$.

In our simulations, we simulate up to $L=20$ layers for the heavy-hex topology and up to $L=15$ layers for the Willow topology. With $\chi=50$ the heavy-hex TNS achieves $f = 0.997$ fidelity (the wavefunction takes up 96MB of RAM -- using double precision complex numbers in each tensor), and with $\chi=20$ the Willow topology TNS achieves $f = 0.859$ fidelity (247MB).
With moderate resources (a single GPU with upwards of 16GB RAM and around $20$ minutes of walltime)
the heavy-hex TNS simulation can be pushed to $\chi \sim 300$ and the Willow TNS to $\chi \sim 50$ for the desired number of layers, which result in a wavefunction memory-cost of 8GB or 13GB, respectively. At this size, the bond dimension of the Willow tensor network is notably larger than that typically considered in literature for square-lattice tensor networks, and extracting accurate information from it, with current methods, beyond the belief propagation approximation is very challenging. The same is not true for the heavy-hex topology for two crucial reasons: \textbf{i)} the lower connectivity means the boundary MPS contraction scheme scales  more favorably in $\chi$ and thus it is cheaper to correct belief propagation with MPS of dimensions $R > 1$ and \textbf{ii)} the much larger loops means smaller corrections are needed to belief propagation, which can already be remarkably accurate for direct measurement of observables.

Figure~\ref{fig:heisenberg-sampling} shows our results in sampling from the evolved TNS with varying boundary-MPS sampling dimensions $R$ (recall we set $R_{x}, R_{n} = R$). We refer to Fig.~\ref{fig:all_heisenerg_poverqs} in the SM for $p(x)/q(x)$ histograms for a wider range of $R$ values. It is clear in all cases that increasing $R$ yields increasingly high quality samples: both the sample KLD decreases and the rate of correct magnetization samples approaches $100\%$. The results in Fig.~\ref{fig:heisenberg-sampling} show that for the hardest Willow TNS increasing from $R =1$ to $R = 20$ provides a dramatic improvement in how well the distribution $q(x)$ reflects the exact encoded probabilities in this 105-qubit quantum state --- although there is still a significant deviation between the two distributions which requires larger $R$ to correct.
Notably, the heavy-hex bond dimension here is $\chi = 50$ and $R \ll \chi$ is sufficient to get the sample KLD below $1$.

\begin{figure}
    \centering
    \includegraphics[width=1\linewidth]{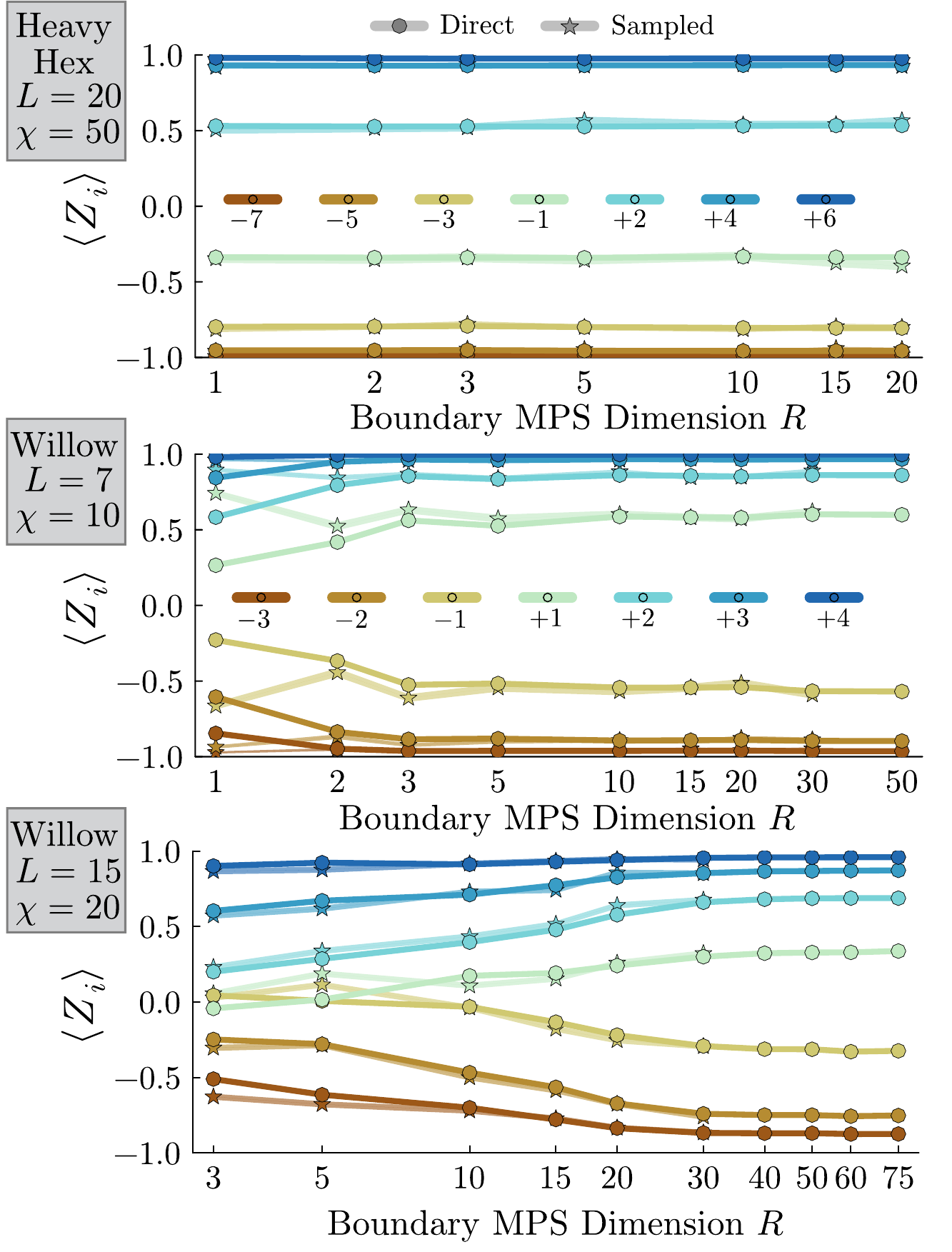}
    \caption{\textbf{Expectation values via direct contraction and sampling for the Heisenberg model.} 
    We show Pauli $Z$ expectation values for the quench dynamics (see Eq. (\ref{eq:HeisenbergProp})) of an initial domain wall after $L$ layers of the Heisenberg model as a function of the boundary MPS dimension $R$. The sites are located on the central row of the lattice and at increasing distance (in terms of number of bonds) from the initial domain wall of the 164-qubit heavy-hex and 105-qubit Willow topologies. The expectation values are estimated both directly through boundary MPS contraction of $\langle \psi \vert \psi \rangle$ and via $1000$ samples generated at the designated boundary MPS dimension (1 std error of the mean is shaded). The tensor network state $\vert \psi \rangle$ has the labeled bond dimension $\chi$. It is clear that the Willow topology generates significantly stronger loop correlations than the heavy-hex topology, even at a third of the circuit depth. Interestingly, at sample KLD values of approximately 2 (c.f., Fig.~\ref{fig:heisenberg-sampling}), the samples can accurately recover local expectation values while the full distribution still differs from the true one. 
    }
    \label{fig:heisenberg_expectations}
\end{figure}

Whilst the sample KLD provides an accurate indicator of the quality of our samples, due to its global nature it can be an overly conservative estimate of accuracy when the desired measurement outcomes of the state are low-weight observables such as one or two-site expectation values. In Fig.~\ref{fig:heisenberg_expectations} we showcase the local $Z_i$ expectation values for sites of varying distances from the center of both topologies --- computed both from the sample distributions illustrated in Fig.~\ref{fig:heisenberg-sampling} and from direct computation using MPS messages to contract the $\langle \psi | Z_i| \psi \rangle$ and $\langle \psi \vert \psi \rangle$.
The heavy-hex topology shows an almost complete insensitivity to the single-site expectation with the boundary MPS dimension both when sampling and directly computing an observable. 
The Willow topology is notably different, even at shallower circuit depths. At $7$ layers, expectation values can be converged with our MPS algorithms but require dimensions $R$ on the order of the bond dimension $\chi$ of the state. Meanwhile, at $15$ layers it becomes much more computationally expensive to converge the single-site expectation value with either direct computation or sample-based computation. Convergence in $R$ is particularly slow closest to the initial domain wall. Here the most entanglement has been generated due to those sites being within the light cone of the simulation for the longest (non-trivial dynamics is only induced on qubits once the light cone reaches them from the domain wall). After $7$ and $14$ layers of the circuit in the Willow and heavy-hex cases respectively, all qubits are within this light-cone.

At this depth, we find we have to go to MPS dimensions $R \sim 75$ to obtain convergence in local expectation values when contracting the norm of this 2D tensor network $\vert \psi \rangle$ which has bond dimension $\chi = 20$. These results thus necessitated the accurate contraction of the norm of a 2D PEPS of very high bond dimension. We achieve this here because, as we systematically increase boundary MPS rank, the workload of our fitting method is increasingly dominated by tensor contractions and allows us to leverage GPU hardware to its fullest extent.

\begin{figure*}
    \centering
    \includegraphics[width=1\linewidth]{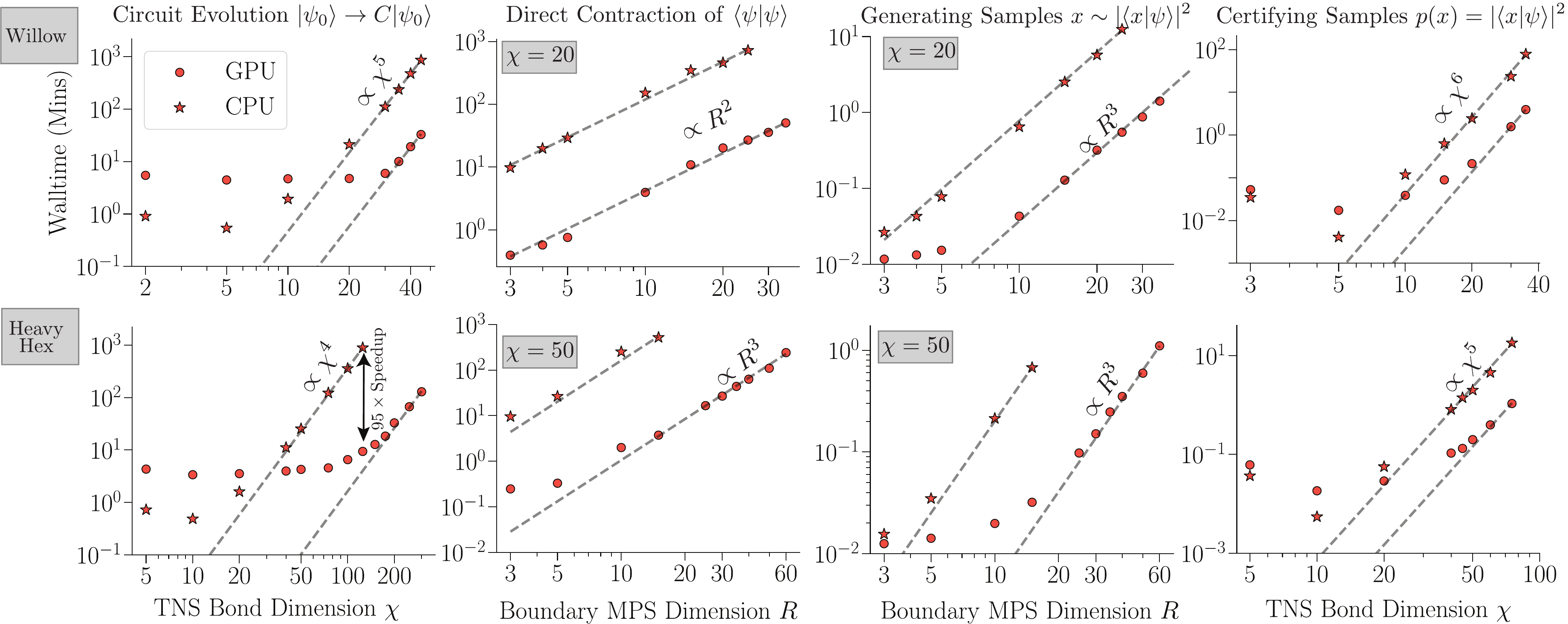}
    \caption{\textbf{Walltimes for evolving, contracting and sampling a $105$-qubit Willow tensor network (top panels) or a $164$-qubit heavy-hex tensor network (bottom panels) respectively}. The tensor network state $\vert \psi \rangle$ is that obtained after $15$ and $20$ layers of the propagator in Eq.~\eqref{eq:HeisenbergProp} with $\delta t J = 0.1$. \textbf{Left)} Time to evolve the tensor network under the propagator, setting a maximum bond dimension $\chi$. \textbf{Center Left)} Time to contract the norm network $\langle \psi \vert \psi \rangle$ with boundary MPS of dimension $R$. \textbf{Center Right)} Average time to generate a sample $x$ from $\vert \psi \rangle$ using a boundary MPS of dimension $R$, following a pre-computed contraction of the norm network $\langle \psi \vert \psi \rangle$. \textbf{Right)} Average time to compute $\langle x \vert \psi \rangle$ given a sample $x$, using a boundary MPS dimension $R = 2\chi$. CPU hardware corresponds to an Intel Xeon 6244 Gold CPU whilst GPU hardware corresponds to an Nvidia RTX A6000. Grey dashed lines show a fit with the annotated polynomial scaling laws, consistent with Table \ref{Table:ComplexityTable}.}
    \label{fig:walltimes}
\end{figure*}

We show a comparison of the relevant walltimes in Fig. \ref{fig:walltimes} on both CPU and GPU for all aspects of the Heisenberg model simulations --- circuit evolution, direct contraction of $\langle \psi \vert \psi \rangle$, sampling from $q(x)$ and verifying samples by calculating $p(x)$. We identify enormous speedups across the board using GPUs which can reach upwards of $90$x even when comparing to multi-threaded CPU calculations. The scaling we observe is consistent with the theoretical scaling in Table \ref{Table:ComplexityTable} for sufficiently large $\chi$ or $R$, although we interestingly note that when directly contracting $\langle \psi \vert \psi \rangle$ the $R^{2}$ scaling term dominates the Willow contraction at the given bond dimension whilst the $R^{3}$ dominates the Heavy-Hex contraction.

\begin{figure}
    \centering
    \includegraphics[width=1\linewidth]{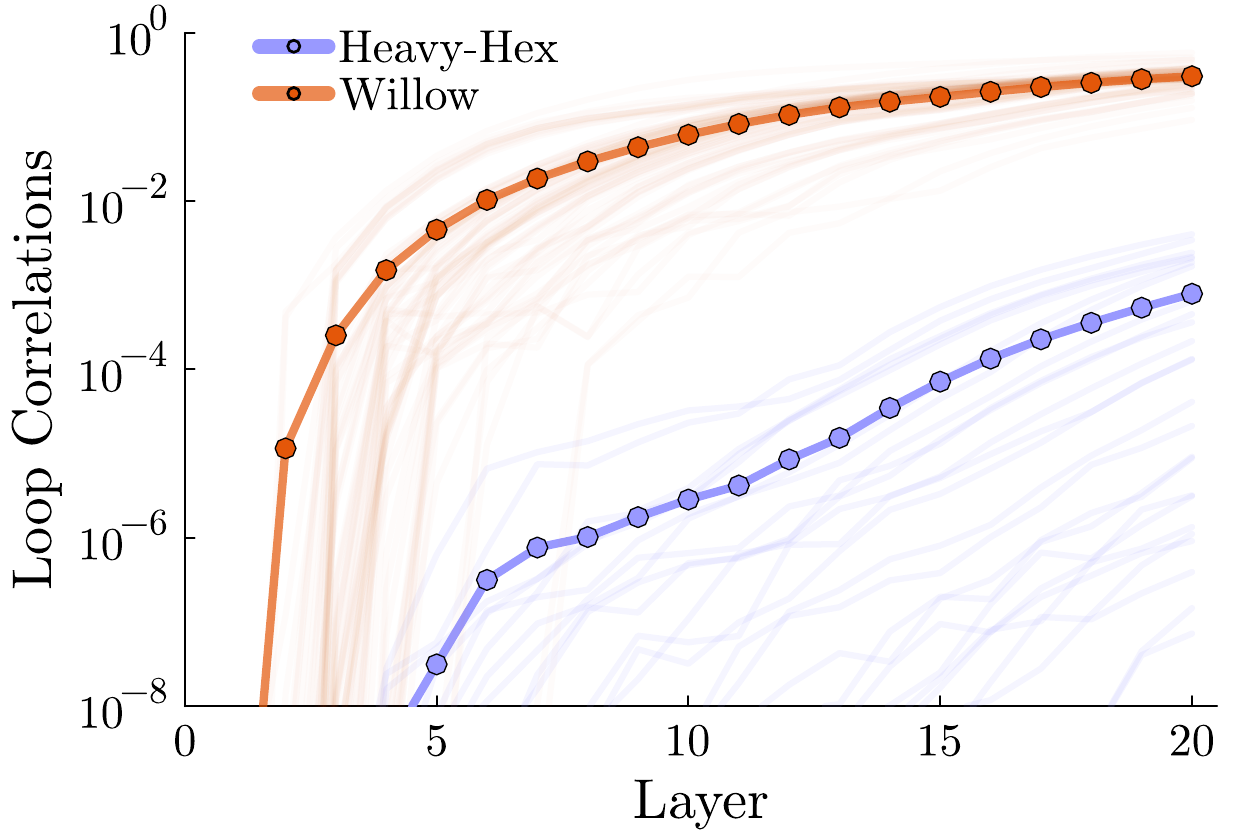}
    \caption{\textbf{Quantifying loop correlations in the heavy-hex and Willow topologies.} We depict all individual approximate BP errors $\varepsilon_l$ per loop in faint colors and their average $\varepsilon$ (see Eqs.~\eqref{eq:avBPError}~\&~\eqref{eq:loopBPerror}) for the discrete-time Heisenberg evolution as a function of the number of Trotter layers. With a loop size of 12, significant loop correlations in the heavy-hex topology can only start arising at 6 layers, with artifacts due to the Trotterization of non-commuting gates arising earlier. In contrast, the Willow topology at that point already exhibits significant loop correlations requiring, e.g., boundary MPS approaches with $R>1$ to extract accurate properties from the states. The large spread between the individual loop errors is explained by the initial location of the spin domain wall, which causes loops near the middle of the lattice to exhibit strong correlations earlier.}
    \label{fig:loopcorrelation}
\end{figure}

Finally, we study the build up of loop correlations from the Heisenberg dynamics. When these correlations are large, they necessitate the aforementioned contraction of $\langle \psi \vert \psi \rangle$ with a large boundary MPS dimension $R$. We compute the first-order approximation to the BP error in both setups (see Eq.~\eqref{eq:avBPError}) as a function of the number of Trotter layers. This error can be seen as quantifying the strength of these loop correlations in the TNS. The results 
are shown in Fig.~\ref{fig:loopcorrelation}. We observe a drastically larger BP error (many orders of magnitude) for the Willow square-lattice versus the heavy-hexagonal lattice. The loops in the heavy-hex lattice are three times as large and thus generically one could expect $\varepsilon ({\rm Heavy \ Hex}) \sim \varepsilon^{3} ({\rm Willow})$ ($0 \leq \varepsilon \leq 1$) based on the exponential scaling of the eigenvalue gap of increasingly long sequences of matrices. The difference we see in Fig.~\ref{fig:loopcorrelation}, however, appears to go beyond even this, and is reinforced by the remarkably accurate BP results numerically observed in Fig.~\ref{fig:heisenberg_expectations} and in Ref.~\cite{tindall2023efficient} for Ising model dynamics on large heavy-hex geometries. These results point to some level of ``loop interference'' in large lattice systems which compounds with the increased loop sizes relative to the Willow topology.

\section{Conclusion}
In this work, we showcased systematically improvable techniques for simulating 2D quantum circuits and their outcomes with planar tensor networks in a scalable, controllable manner. These classical networks make a natural ansatz for simulating upcoming quantum computers, including quantum circuits running on superconducting processors.
 
We applied gates in the circuit to the tensor network via the belief propagation-based simple update procedure, whilst sampling was performed using generalized Matrix Product State and Matrix Product Operator contraction routines, which allow the approximate contraction of arbitrary planar tensor networks. Importantly, we identified reliable metrics for both the fidelity of the tensor network and the quality of the samples in order to attest to the quality of our simulations. 

By applying these techniques to the local unitary Jastrow ansatz (LUCJ) circuits introduced in Ref.~\cite{robledomoreno2025}, we generated samples which are drawn, to numerical precision, from the exact underlying distribution. We also showed the generality of our methods, simulating, on moderate timescales, the highly-entangling quench dynamics of a domain wall in the two-dimensional Heisenberg model on both IBM's heavy-hex processor architecture and Google's latest Willow processor. We exploited the potential for GPU speedup latent in these contraction methods, to accurately sample and contract the norm of 2D Tensor Networks of very large bond dimension.

Crucially, our results demonstrated that the loop correlations generated are remarkably low for quantum circuits realized on heavy-hexagonal processors, which leads to procedures with boundary MPS of very low dimension (and consequently belief propagation) yielding highly accurate samples and expectation values even at large circuit depths. In fact, for local observables, we observe immediate convergence with $R=1$ contraction procedure in all cases, implying that significantly deeper circuits or larger time steps are required for classical hardness. The same is not true for the Willow processor, whose topology means that extracting accurate expectation values from moderate-depth circuits (such as those encoding the time dynamics of the Heisenberg model) can require significant computational resources.

It should be pointed out that there are certain, finite dimension, pathological tensor network states one can construct for which perfect sampling must require resources (the boundary MPS dimension $R$) growing exponentially in the system size \cite{verstraetepepshard2006}. We do not observe signatures of such states here, with the value of $R$ necessary to obtain accurate results in our finite size instances well below the largest value it can take, which can be estimated from the bond dimension as $\chi^{\sqrt{N_{\rm qubits}}}$ for the states in this work. This is likely because they are generated from circuits which encode local, physical interactions and thus there is a finite velocity associated with information spreading in the system. 

Our results here highlight how geometry plays a crucial role in the complexity of quantum circuits and provide a state-of-the-art framework --- with corresponding open-source code with both CPU and GPU support \cite{TensorNetworkQuantumSimulator} -- for simulating quantum circuits with planar tensor networks. We hope that these tensor networks and the underlying classical simulation methods in general become more widely used by those at the forefront of developing quantum devices and their applications. 

\section*{Declarations}

\subsection*{Acknowledgements}
The authors acknowledge helpful discussions with Zo\"e Holmes, Miles Stoudenmire, and Matt Fishman. The authors would like to thank Miles Stoudenmire, Supanut Thanasilp and Haochen Li for feedback on this manuscript. The authors also acknowledge Kevin Sung and Javier Robledo-Moreno for discussions and providing the scripts necessary to generate the LUCJ circuits.

\subsection*{Competing Interests}
The authors declare no competing interests.

\subsection*{Funding}
MSR is grateful for support by the Flatiron Institute during the Pre-Doctoral Researcher program, and JT is grateful for ongoing support through
the Flatiron Institute, a division of the Simons Foundation.
MSR acknowledges funding from the 2024 Google PhD Fellowship and the Swiss National Science Foundation [grant number 200021-219329].

\subsection*{Code Availability}
The documented, open source Julia code for reproducing the results in this work is available at TensorNetworkQuantumSimulator.jl \cite{TensorNetworkQuantumSimulator}, an open source wrapper --- built off of ITensors.jl \cite{fishman2022itensor} --- for simulating quantum circuits with tensor networks of arbitrary topology.
\newline

\subsection*{Data Availability}
All study data are included in this article and its Supplementary Materials.

\subsection*{Contributions}
M.S.R and J.T wrote the manuscript.
M.S.R and J.T prepared the figures and ran the simulations.
J.T and M.S.R wrote the software and algorithms necessary to produce the results.

\clearpage

\onecolumngrid

\section*{Supplementary Material for ``Simulating and Sampling from Quantum Circuits with 2D Tensor Networks"}
\date{\today}

\renewcommand{\thefigure}{S\arabic{figure}}
\renewcommand{\thetable}{S\arabic{table}}
\renewcommand{\theequation}{S\arabic{equation}}

\setcounter{figure}{0}
\setcounter{table}{0}
\setcounter{equation}{0}

\subsection*{Sampling from a Tensor Network State}

Here we detail, in-depth, a computational procedure to sample from a planar tensor network representation of the wavefunction $\vert \psi \rangle$. Our method and implementation can be seen as a generalization of the algorithm detailed in Ref.~\cite{Vieijra2021sampling} to arbitrary planar topologies.

\begin{figure*}
    \centering
    \includegraphics[width=1\linewidth]{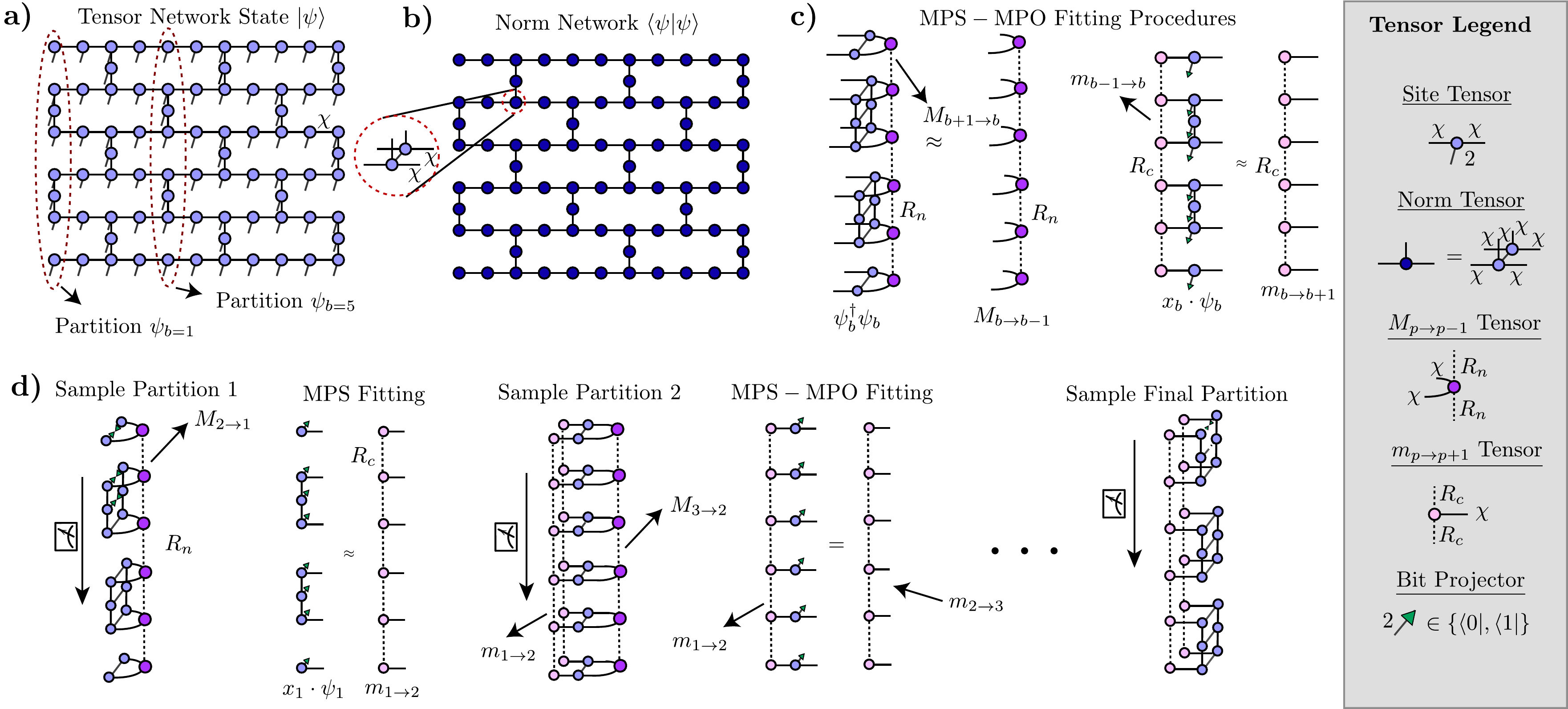}
    \caption{\textbf{Sampling from planar Tensor Network States.} a) Planar tensor network representation of a wavefunction $\vert \psi \rangle$ with bond dimension $\chi$. A heavy-hex topology is chosen for illustrative purposes. The tensors in the network are grouped into partitions $b = 1, 2 \hdots N_{b}$ - with a column-based partition illustrated here. b) Norm network $\langle \psi \vert \psi \rangle$ with the individual nodes formed from uncontracted pairs of tensors in  $\vert \psi \rangle$. c) MPS-MPO fitting procedures that form the workhorse of the sampling procedure. The contraction of incident MPS with either the MPO defined by the two-layer partition $\psi^{\dagger}_{b}\psi_{b}$ or the MPO defined by the single-layer partition $x_{b} \cdot \psi_{b}$ are approximated with an outgoing MPS via a variational one-site fitting procedure that works on MPOs of arbitrary structure. Other procedures can be used, but the one-site fitting procedure scales most favorably in bond dimension $\chi$.
    d) Procedure for generating a single bitstring $x$ assuming the pre-computation of the set of MPS $\{M_{N_{b} \rightarrow N_{b} - 1},  \hdots M_{3 \rightarrow 2},M_{2 \rightarrow 1} \}$ via the MPS-MPO fitting procedure on the two-layer tensor network state. The first partition $\psi^{\dagger}_{1} \psi_{1}$ is sampled with the MPS $M_{2 \rightarrow 1}$ incident. This is done by contracting the structure from top to bottom and sequentially splitting open the bonds connecting the bra and ket tensors to form the one-site reduced density matrix and sample from it, conditioned on the already sampled sites above it. The structure $\langle x_{1} \vert \psi_{1} \rangle$ is then fit to an MPS $m_{1 \rightarrow 2} $ and the second partition is sampled. The structure $m_{1 \rightarrow 2}  \cdot x_{2} \cdot \psi_{2}$ is then fit to an MPS $m_{2 \rightarrow 3}$ and the next partition sampled. This is repeated until all partitions are sampled, yielding a bitstring $x$ from the distribution $q(x)$ defined by the dimension $R_{n}$ and $R_{x}$ chosen for the MPS $M_{i + 1 \rightarrow i}$ and $m_{i \rightarrow i + 1}$ respectively. Computation of $p(x) = \vert \langle x \vert \psi \rangle \vert^{2}$ can be done either by selecting a sufficiently large $R_{x}$ or a separate contraction of the network $\langle x \vert \psi \rangle$. A legend is included to show the different tensors which appear and the dimension of their respective indices.}
    \label{fig:tns-sampling}
\end{figure*}

Consider a planar tensor network $\vert \psi \rangle$ which we wish to draw samples $x$ from. In Fig.~\ref{fig:tns-sampling} we illustrate this with the example of a network with heavy-hex topology. The tensors of the network are first grouped into partitions $b = 1, 2, \dots N_{b}$ such that the topology of the network following this partitioning is a single line with edges between sequential partitions. This is most naturally achieved by partitioning the network by its rows or columns (we show the choice of column partitions in Fig.~\ref{fig:tns-sampling}), although other choices are possible. We define the subset of tensors in a given partition with $\psi_{b}$. We also define the norm network $\langle \psi \vert \psi \rangle$ and an identical partitioning such that the partitions $T_b = \psi^{\dagger}_{b}\psi_{b} $ are formed from the tensors in $\psi_{b}$ and their conjugates. Those $T_b$ are generally MPOs, apart from the first and last partition at the boundaries, where they are MPS.

A crucial ingredient of the sampling procedure is a generalized MPS-MPO and MPS fitting method. Specifically, we need to be able to approximate the contraction of a Matrix Product State $M_{b +1 \rightarrow b}$ with a partition, i.e. $M_{b +1 \rightarrow b} \cdot T_b $ with another MPS $M_{b \rightarrow b -1}$, and also approximate the contraction of a Matrix Product State $m_{b - 1 \rightarrow b}$ with a sampled partition, i.e. $m_{b - 1 \rightarrow b } \cdot X_b$ with another MPS $m_{b \rightarrow b + 1}$, where $X_b = x_{b} \cdot \psi_{b}$ are again generally MPOs apart from the boundaries. This process is illustrated in Fig.~\ref{fig:tns-sampling}c. The fitting can be achieved most efficiently by a one-site variational fitting procedure~\cite{verstraete2008matrix} which forms an initial guess for the output MPS and approximately maximizes the overlap between the output MPS and the MPS-MPO contraction by variationally sweeping through the tensors of the output MPS and replacing them with the derivative of the MPS-MPO-MPS contraction with respect to that tensor. We have implemented this fitting procedure algorithmically in a highly generic, efficient way, such that incoming MPS can be fit to MPOs of arbitrary structure  (i.e. the MPOs can just be any tensor network which maps an MPS to another MPS).

It is possible to obtain an error in this variational MPS-MPO fitting by keeping track of
\begin{equation}
    f = \frac{\vert \langle\widetilde{\text{MPS}}|\widehat{\text{MPO}}|\text{MPS}\rangle \vert^{2}}{\langle \text{MPS} \vert \text{MPS} \rangle \langle\widetilde{\text{MPS}}|\widehat{\text{MPO}}|\widehat{\text{MPO}}|\widetilde{\text{MPS}}\rangle},
\end{equation}
during the procedure, which is bounded between $0$ and $1$. This quantity has the issue, however, in needing to calculate the norm of the state being fitted, i.e $\langle\widetilde{\text{MPS}}|\widehat{\text{MPO}}|\widehat{\text{MPO}}|\widetilde{\text{MPS}}\rangle$ which is more computationally expensive than the fitting itself (it will scale rather severely in $\chi$ and $R$). It is often not even possible to precisely calculate it during simulations such as the ones we do in the paper as these involve substantial values of $R$ and $\chi$. 

\par In order to determine convergence in our fitting procedure we monitor the reduced quantity 
\begin{equation}
    \tilde{f} = \frac{\vert \langle\widetilde{\text{MPS}}|\widehat{\text{MPO}}|\text{MPS}\rangle \vert^{2}}{\langle \text{MPS} \vert \text{MPS} \rangle},
\end{equation}
which can be kept track of without any extra computation. This is no longer a proper error metric but clearly converges in exactly same way as $f$ as it is just related by a proportionality constant $1/\langle\widetilde{\text{MPS}}|\widehat{\text{MPO}}|\widehat{\text{MPO}}|\widetilde{\text{MPS}}\rangle$ which is fixed. Thus, we rely on convergence in properties (observables, norms) in $R$ in order to assess the accuracy/ errors in our MPS-MPO fitting method. 

With the fitting procedure in hand, the sampling procedure can be defined as illustrated in Fig. \ref{fig:tns-sampling}d. First, the last partition $T_{N_{b}}$ is approximated by an MPS $M_{N_{b} \rightarrow N_{b} - 1}$ of dimension $R_{n}$. Then the MPS-MPO contraction $M_{N_{b} \rightarrow N_{b} - 1} \cdot T_{N_{b}-1}$ is approximated with an MPS $M_{N_{b} - 1 \rightarrow N_{b} - 2}$. This last procedure is then repeated for the partitions $b = N_{b} - 2$ through $b = 2$ yielding the set of MPS $\{M_{N_{b} \rightarrow N_{b} - 1},  \hdots M_{3 \rightarrow 2},M_{2 \rightarrow 1} \}$. This procedure only needs to be done once, independent of the number of samples one wishes to draw.

\begin{figure*}
    \centering
    \includegraphics[width=0.95\linewidth]{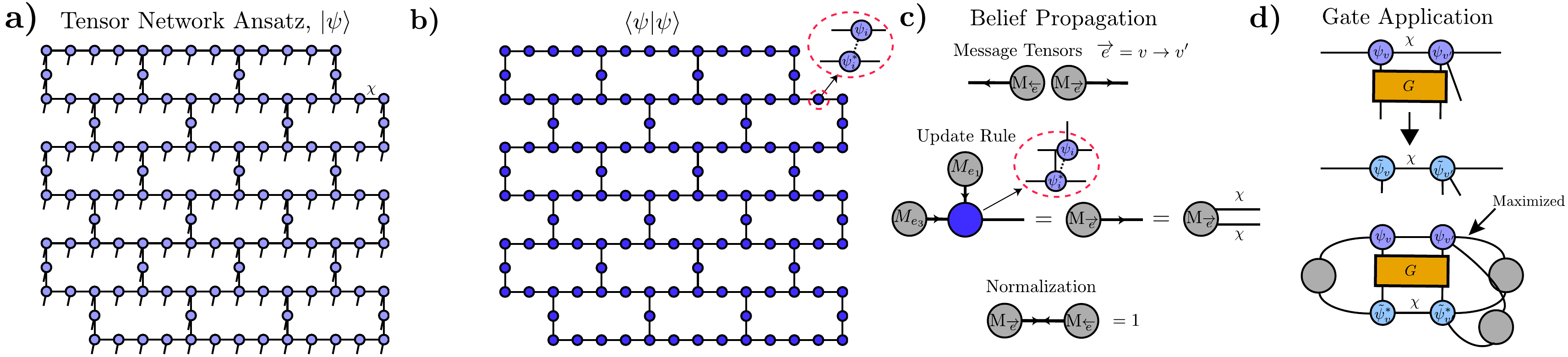}
    \caption{\textbf{Belief propagation algorithm and gate application for a Tensor Network State.} a) Planar tensor network representation of a wavefunction $\vert \psi \rangle$ with bond dimension $\chi$. A heavy-hex topology is chosen for illustrative purposes.  b) Norm network $\langle \psi \vert \psi \rangle$ with the individual nodes formed from uncontracted pairs of tensors in  $\vert \psi \rangle$. c) Belief propagation algorithm. Message tensors are initialized in each direction on every edge of the norm network and self-consistently updated until convergence subject to a normalization condition.
    d) Gate application. A gate is applied to a pair of sites $v$ and $v'$ conditioned on the BP approximation by updating the corresponding tensors $\psi_{v}$ and $\psi_{v'}$ with those that maximise their overlap with the original tensors, the gate and the incoming message tensors to that region.}
    \label{fig:bpgateapplication}
\end{figure*}

Next, for each sample desired, the first partition $T_{1}$ is sampled conditioned on the incident MPS $M_{2 \rightarrow 1}$. This can be done by moving through the partition, qubit by qubit, and forming the one-site reduced density matrix conditioned on the incident MPS and any qubits already sampled in the partition. The result is a sample $x_{1}$ of all qubits in the partition conditioned on $M_{2 \rightarrow 1}$ as an approximation of the contraction of the rest of the network. Next, the tensors in $X_{1}$ are fit to a MPS $m_{1 \rightarrow 2}$ of bond dimension $R_{x}$.
The partition $b = 2$ can then be sampled conditioned on the incident MPS $m_{1 \rightarrow 2}$, $m^{\dagger}_{1 \rightarrow 2}$ and $M_{3 \rightarrow 2}$.  Then, the contraction of $m_{1 \rightarrow 2}$ with $ x_{2} \cdot \psi_{2}$ is fit to a MPS $m_{2 \rightarrow 3}$ of maximum bond dimension $R_{x}$ and the partition $b = 3$ is sampled. This procedure is repeated until all columns are sampled yielding the bitstring $x = x_{1}x_{2} \hdots x_{N_{b}}$ in a manner which scales linearly with the number of qubits. 

The resulting bitstring is drawn from the distribution $q(x)$ defined by the selected MPS bond dimensions $R_{x}$ and $R_{n}$, and \textit{not necessarily} from the actual distribution $p(x) = \vert \langle x \vert \psi \rangle \vert^{2}$ of the tensor network state. Equality is only achieved if $R_{x}$ and $R_{n}$ are large enough such that there is no error in the fitting procedures.
The probability $q(x)$ is returned immediately from the sampling procedure as it is just the product of the individual probabilities $q(x_{q})$ when sampling the reduced density matrix for each qubit $q$. The probability $p(x)$ can be obtained in one of two ways: if the MPS dimension $R_{x}$ used is large enough such that only minimal truncations are made in the fitting procedures for the $m_{i \rightarrow i + 1}$ then it is the square of the MPS-MPS contraction $m_{N_{b} - 1 \rightarrow N_{b}} \cdot X_{N_{b}}$. If significant truncations are made then it can be obtained independently via contraction of the planar tensor network $\langle x \vert \psi \rangle$ with either sequential MPS contractions or a separate method such as loop corrections. Notably, this separate verification step scales better than sampling with a higher boundary MPS dimension and is not strictly necessary if only the samples are desired.

The set of ratios $ \{ \frac{p(x_{1})}{q(x_{1})}, \frac{p(x_{2})}{q(x_{2})}, \frac{p(x_{3})}{q(x_{3})} \hdots \frac{p(x_{m})}{q(x_{m})} \}$ for a series of $m$ samples provides clear information about the quality of the samples generated for the chosen $R_{x}$ and $R_{n}$. Moreover, they can be used to correct the computation of an observable from those samples via the ``importance sampling" formula

\begin{equation}
    \langle \psi \vert O \vert \psi \rangle \approx \frac{1}{N}\sum_{i= 1}^{m}\frac{p(x_{i})}{q(x_{i})}\langle x_{i} \vert O \vert x_{i} \rangle
\end{equation}
with $N  = \frac{1}{m}\sum_{i}\frac{p(x_{i})}{q(x_{i})}$ an approximation for the norm of the wavefunction. This approximation becomes equality in the limit $m \rightarrow \infty$ and all $\frac{p(x_{i})}{q(x_{i})}$ are finite. 

\textit{Computational Complexity} - In the following and throughout this work, we take $R_{x} = R$ and $R_{n} = R$. The complexity of generating samples $x$ is highly dependent on the coordination number of the tensor network. For $R \leq \chi$, with $\chi$ the bond dimension of the tensor network, then on a (rotated or unrotated) square lattice processor with $z=4$, such as the Willow processor, and a total number of qubits or tensors $N_{\rm qubits}$, $m$ samples can be obtained with time complexity $\mathcal{O}(N_{\rm qubits}\chi^{5} R^{3}) + \mathcal{O}(m N_{\rm qubits}\chi^{4} R^{3})$ upon partitioning the network by either its columns or rows.
Meanwhile on a heavy-hex architecture where $z=3$, $m$ samples can be obtained with time complexity $\mathcal{O}(N_{\rm qubits}\chi^{4} R^{3}) + \mathcal{O}(m N_{\rm qubits}\chi^{3} R^{3})$ upon partitioning the network by either its columns or rows.

\subsection*{Applying Gates to a Tensor Network State}

In this work, we apply gates to our tensor network ansatz for the many-body wavefunction $\vert \psi \rangle$ using message tensors obtained from the belief propagation algorithm. Specifically, given a tensor network representation of $\vert \psi \rangle$, we form the network $\langle \psi \vert \psi \rangle$ from two copies of the tensor network. We group the individual tensors $\psi_{v}$ and their conjugates $\psi_{v}^{*}$ together, such that the norm network has the same structure as the original network. This is illustrated in Fig.~\ref{fig:bpgateapplication}b. It is crucial for efficiency that this grouping of tensors remains a book-keeping operation and the individual tensors are not contracted. We then initialize \textit{message tensors} in both directions on each edge $v \leftrightarrow v'$ of the network. As such, these message tensors each possess the virtual indices corresponding to the grouped pair of tensors grouped ($\psi_{v}$, $\psi^{*}_{v}$ and $\psi_{v'}$, $\psi^{*}_{v'}$) at each end of the edge. A self-consistent update rule for the message tensors is defined, with the message tensor on an edge $v \rightarrow v'$ equal to the incoming message tensors to $v$ (excluding the message from $v'$ to $v$) multiplied by the local tensors  $\psi_{v}$ and $\psi^{*}_{v'}$. Imposing the normalization condition that the norm of a message tensor is $1$ this update rule can be iterated until appropriate convergence of all message tensors~\cite{Alkabetz2021, tindall2023gauging, evenbly2025loopseriesexpansionstensor}. These message tensor details are outlined in Fig.~\ref{fig:bpgateapplication}c.

\begin{figure}
    \centering
    \includegraphics[width=0.5\columnwidth]{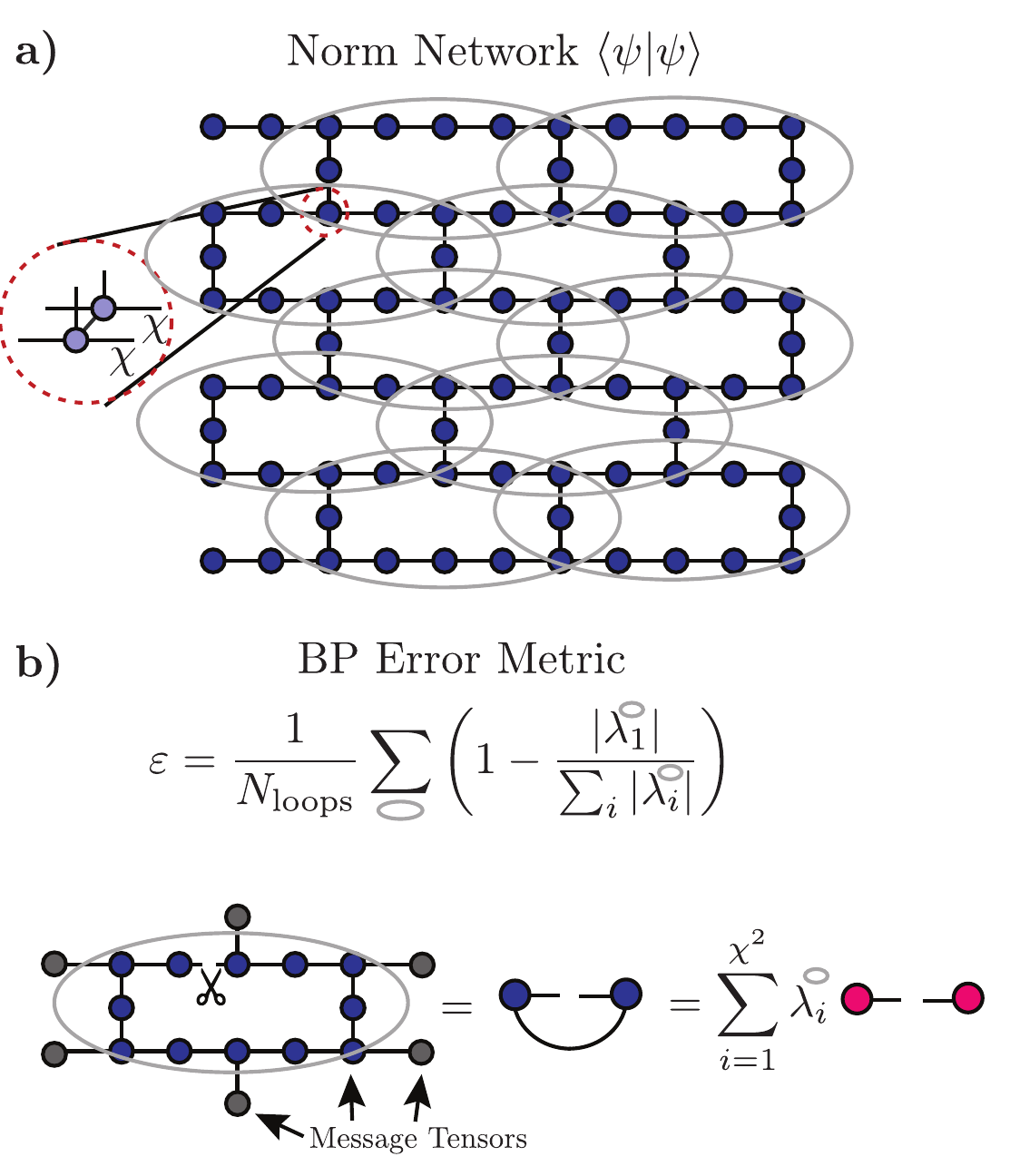}
    \caption{\textbf{Computation of a belief propagation error metric.} a) Norm network $\langle \psi \vert \psi \rangle$ associated with a heavy-hex tensor network representation of $\vert \psi \rangle$ with the individual nodes formed from uncontracted pairs of tensors in  $\vert \psi \rangle$. The primitive loops (set of smallest loops) of the lattice are ringed. b) An error metric (see Eq.~\eqref{eq:avBPError}) can be defined by averaging the separability index for each loop. The separability index is computed from the ordered (by absolute value) eigenvalues of the transfer matrix formed when inserting BP message tensors on the boundary of the loop and splitting an edge of the loop open.}
    \label{fig:bperror}
\end{figure}

These messages can then be used to condition the singular value decomposition during the application of a two-site gate to the network. Specifically, when applying a gate to a local pair of sites $v$ and $v'$ in the network, the tensors $\psi_{v}$ and $\psi_{v'}$ are replaced with a new pair of normalized tensors $\tilde{\psi}_{v}$ and $\tilde{\psi}_{v'}$ sharing a bond of specified dimension $\chi$ such that they maximise the overlap 
\begin{equation}
    C = G \cdot \psi_{v} \cdot \psi_{v'
    } \cdot \tilde{\psi}^{*}_{v} \cdot \tilde{\psi}^{*}_{v'} \cdot \prod_{e}M_{e}
\end{equation}
where $\prod_{e}M_{e}$ is the product of all messages along the edges incident to the region consisting of $v$ and $v'
$. This quantity is illustrated in Fig. \ref{fig:bpgateapplication}d and the tensors can be identified by gauging the region with the square root of the incoming message tensors, applying the gate, performing a singular value decomposition, and ungauging the region with the inverse square root of the incoming message tensors. If the bond dimension $\chi$ is chosen such that no singular values are thrown away, the gate application is exact. More details can be found in Ref.~\cite{tindall2023gauging}.

\subsection*{Computing the BP Error}
As discussed in the main text, an error metric which can be associated when contracting a tensor network via BP is obtainable from the spectrum of eigenvalues $\lambda^{l}_{1}, \lambda^{l}_{2}, \hdots $ of the transfer matrices formed using primitive loops (the set of loops of smallest size) of the tensor network --- see Eq. \eqref{eq:avBPError} for a definition. For the norm network $\langle \psi \vert \psi \rangle$, this spectrum can be obtained exactly in $\mathcal{O}(N_{\rm qubits}\chi^{6} )$ time, whilst it can be reduced to $\mathcal{O}(N_{\rm qubits}\chi^{z+1}  k)$ time with a Krylov-based method if only the $k$ smallest eigenvalues are computed. In Fig.~\ref{fig:bperror} we illustrate the procedure for computing these eigenvalues.

\clearpage

% \section{Additional Numerical Results}
\begin{figure*}
    \centering
    \includegraphics[width=0.6\linewidth]{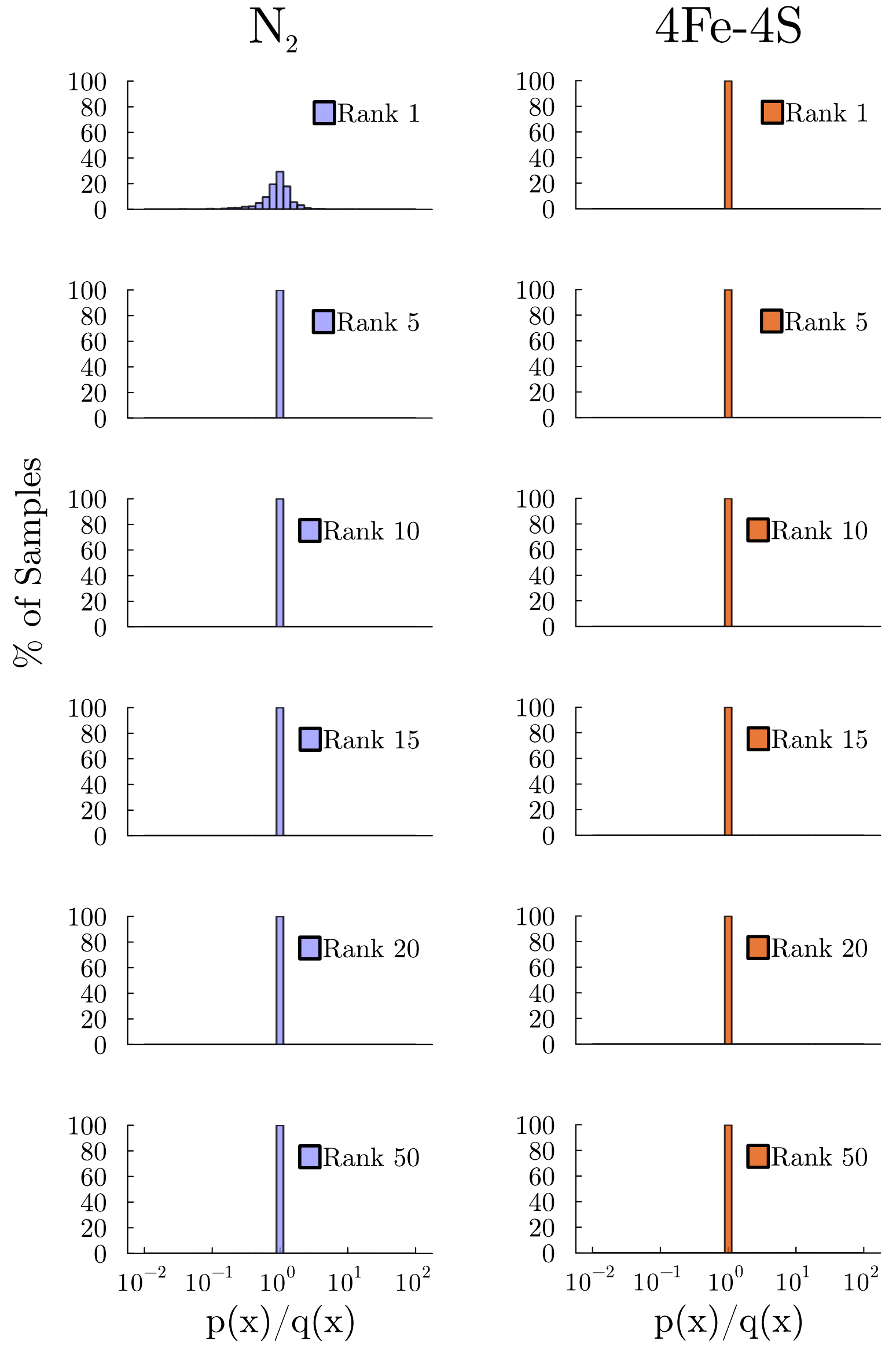}
    \caption{\textbf{Extended sampling probability distributions for local unitary Jastrow circuits.} Histograms of the probability ratios $p(x)/q(x)$ for 1000 samples drawn from the 52-qubit N$_{2}$ (left) and 72-qubit $4$Fe-$4$S (right) molecular simulations across varying boundary MPS dimensions $R$. As discussed in the main text, the $4$Fe-$4$S simulation exhibits effectively no loop correlations, resulting in practically exact sampling (ratios tightly concentrated at $10^0 = 1$) even at the lowest bond dimension $R=1$. In contrast, the N$_{2}$ molecule exhibits a broader distribution at $R=1$ but rapidly converges to generate practically exact samples for $R \ge 5$. These distributions provide a comprehensive view of the convergence behavior summarized in Fig.~\ref{fig:chemistry_sampling}.}
    \label{fig:all_chemistry_poverqs}
\end{figure*}

\begin{figure*}
    \centering
    \includegraphics[width=0.9\linewidth]{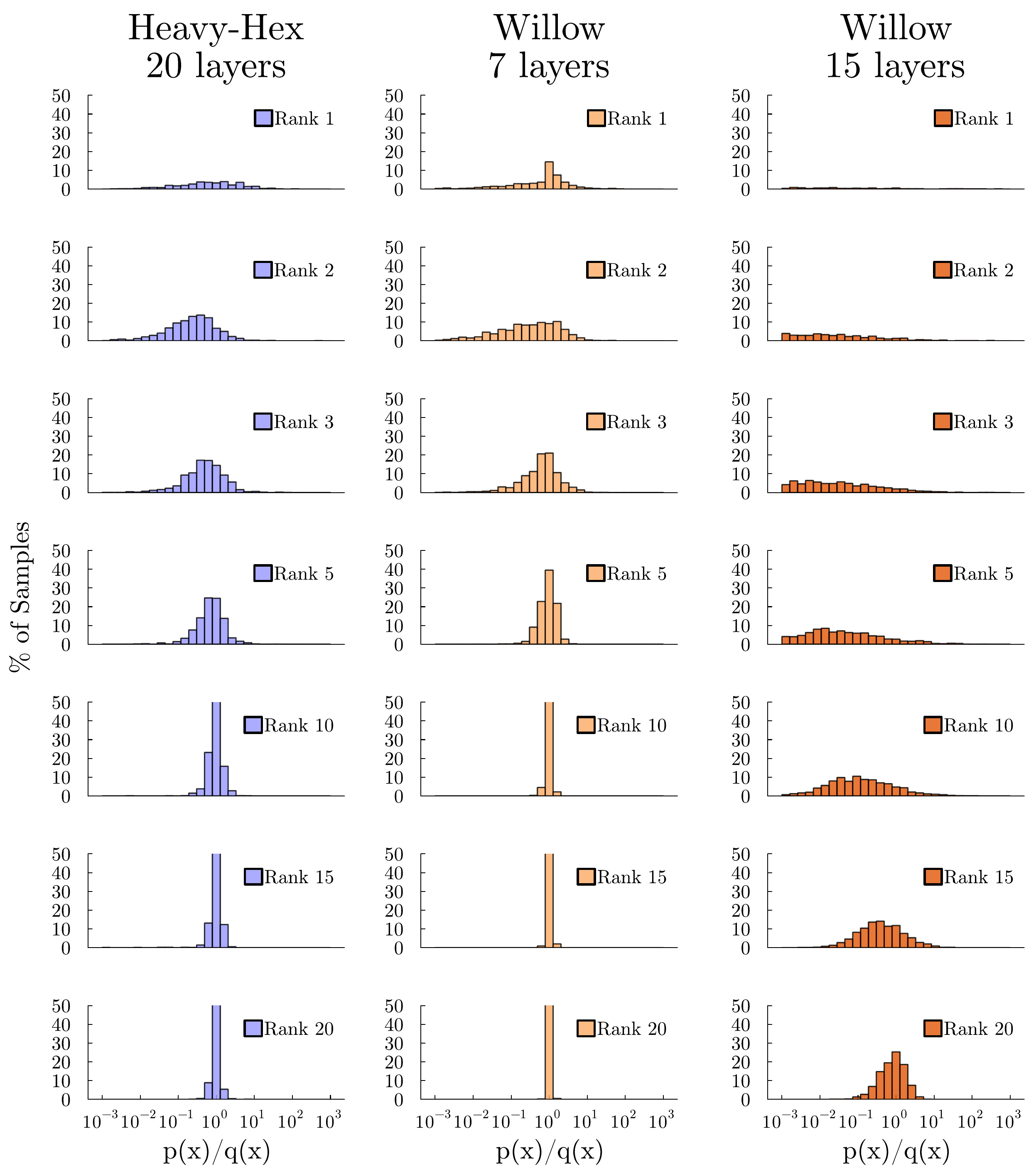}
    \caption{\textbf{Extended sampling probability distributions for the quench dynamics of the Heisenberg model.} Histograms detailing the evolution of $p(x)/q(x)$ probability ratios for 1000 samples as a function of the boundary MPS dimension $R$. The columns correspond to the heavy-hex topology with 20 layers (left) and the Willow topology with 7 layers (middle) and 15 layers (right). For the most complicated 15-layer Willow case, low sampling dimensions ($R \le 3$) yield sample probabilities that are off by many orders of magnitude, causing the sample distributions to fall almost entirely outside the plotted limits. As $R$ increases to $20$, the distributions systematically narrow across all topologies, yielding higher-quality samples with heavily concentrated probability ratios. This figure extends the specific $R=1$ and $R=20$ boundary cases shown in Fig.~\ref{fig:heisenberg-sampling}.}
    \label{fig:all_heisenerg_poverqs}
\end{figure*}

\end{document}